\documentclass[showpacs,twocolumn,aps,prl]{revtex4}
\usepackage{amsmath}
\usepackage{ulem}
\usepackage{amsfonts}
\usepackage{amssymb}
\usepackage{amsthm}
\usepackage{graphicx}

\usepackage{subfigure}
\begin{document}
%----------------------------------------------------------------%
\title{Self-sustained current oscillations in spin-blockaded
quantum dots}
\author{B. Hu$^{1}$ and X.R. Wang$^{2,1}$}
\email{[Corresponding author:]phxwan@ust.hk}
\affiliation{$^{1}$Physics Department, The Hong Kong University of 
Science and Technology, Clear Water Bay, Kowloon, Hong Kong}
\affiliation{$^{2}$School of Physics, Wuhan University, Wuhan, 
P. R. China}
\date{\today}

\begin{abstract}
Self-sustained current oscillation observed in spin-blockaded 
double quantum dots is explained as a consequence of periodic 
motion of dynamically polarized nuclear spins (along a limit 
cycle) in the spin-blockaded regime under an external magnetic 
field and a spin-transfer torque. It is shown, based on the 
Landau-Lifshtz-Gilbert equation, that a sequence of semistable 
limit cycle, Hopf and homoclinic bifurcations occurs as the 
external field is tuned. The divergent period near the homoclinic 
bifurcation explains well why the period in the experiment is so 
long and varies by many orders of magnitudes.
\end{abstract}
\pacs{73.63.Kv, 72.25.Rb, 76.60.-k, 05.45.-a}
% 05.45.-a Nonlinear dynamics and nonlinear dynamical systems

\maketitle
\begin{figure}
  \includegraphics[width=3.5 in]{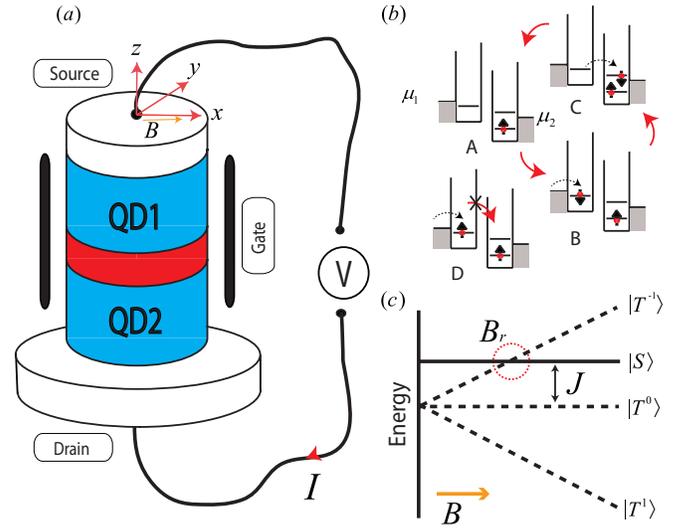}\\
\caption{(Color online) (a) Schematic setup of spin-blockaded double 
quantum dots device.
(b) Four possible configurations in spin-blockaded regime.
One spin-up electron is in the right quantum dot below the chemical
potentials in configuration A. In configuration B, one electron on
each of two dots form a spin singlet state. Configuration C is
the spin singlet state of two electrons in the right dot.
Spin-blockaded configuration D is a spin triplet state of two
dots with one electron on each. (c) Schematic of field-dependence
of triplet and singlet states. Triplet state $|T^{-1}>$ cross with
singlet state $|S>$ near the resonance field $B_r$ (red circle).}
\label{fig1}
\end{figure}

%-----------------------------------------------------------%
Quantum dots, also known as artificial atoms, have many properties 
of natural atoms such as discrete energy levels and shell structures 
\cite{general1,general2}. The physics involved in quantum dots is 
very rich because of tunability and comparability of three energy 
scales: level spacing, Coulomb interaction, and thermal energy. 
Unlike natural atoms, quantum dots allow transport measurements. 
Many interesting transport phenomena like resonant tunnelling, 
Coulomb blockade, spin-blockade, Kondo effects, quantum 
conductance etc. have been observed and explained. 
Applications in nano-electronics, spintronics and quantum computing  
due to possible long coherence time have been proposed and in some 
cases implemented. It is known that both the nuclear and electron spin 
degrees of freedom in semiconductor nanostructures can be manipulated 
by the hyperfine interactions \cite{general3,tarucha1,dobers,tarucha}. 
In the endeavor of using hyperfine interaction to manipulate the 
electron transport in quantum dots, one long-term unexplained 
intriguing phenomenon was observed by Ono and Tarucha in 2004 
\cite{tarucha} in spin-blockaded double quantum dots: tunneling 
current $I$ under a DC-bias, as artistically shown in Fig. \ref{fig1}, 
oscillates with a period up to minutes under certain conditions. 
In this letter, we show that dynamically polarized nuclear spins 
(DPNS), governed by the generalized Landau-Lifshitz-Gilbert equation, 
can oscillate with time under a DC-bias in a magnetic field window. 
The back effect of the DPNS oscillation on the electron tunneling 
leads to the experimentally observed self-sustained current 
oscillation.

A coherent theory should be consistent with the following 
experimental findings: 1) The oscillations observed only in the 
spin-blockaded regime in a magnetic field window accompany by a 
current jump. 2) Both the period and amplitude increase with the
field before the oscillation disappears. 3) The oscillation period 
can be tuned by the magnetic field by several orders of magnitude 
from seconds (the instrumental limit in the experiment) up to 
several minutes. 4) The oscillation is closely related to the 
motion of dynamically polarized nuclear spins (DPNS) 
\cite{tarucha, rudner1,rudner2,inoshita,lopez}. It is clear that 
the oscillation theory for semiconductor superlattices \cite{xrw1} 
is not applicable because the period would be order of 100$ns$ 
electron tunneling time, instead of observed seconds and minutes.
Thermal and impurity effects were also ruled out \cite{tarucha}.
Although the nuclear spins in a quantum dot is known to have a very 
long (up to seconds) relaxation time, the relaxation process is not a
periodic motion so that it cannot be the cause of the oscillation.

In order to construct a sensible model, let us examine the plausible 
microscopic process of electrons and nuclei in spin-blockaded double 
quantum dots. In the experiment \cite{tarucha}, two vertical disk-like 
quantum dots (InGaAs-AlGaAs multilayer structure) are weakly connected 
in series between source and drain (illustrated in Fig. \ref{fig1}a). 
Four possible configurations in the spin-blockaded regime are showed 
in Fig. \ref{fig1}b: One electron is trapped in the right dot 
(configuration A). The second electron, hopping from the source lead 
to the left dot, forms either spin singlet (B) or triplet (D) states. 
Electron tunneling cycle $A\rightarrow B \rightarrow
C\rightarrow A$ is allowed while tunneling is blockaded in D,
here C is spin singlet state of two electrons on the right dot.
As shown in Fig. \ref{fig1}c, three triplet states ($|T^0>, |T^{\pm
1}>$) are degenerated in the absence of an external magnetic field,
and are below the spin singlet state $|S>$ by an energy $J$ \cite{ent},
order of inter-dot exchange energy. Due to the spin blockade
and weak coupling of dots with two leads, a leakage current of order
of 1pA, corresponding to 100$ns$ electron tunneling time, exists.
An external field lifts the triplet degeneracy, and $|S>$
and $|T^{-1}>$ states shall anti-cross around certain field.
The current experiences a jump near the crossing field $B_r$ because
spin blockade is partially removed by spin flipping there, and
transition from configuration D to B becomes possible.
It is known that this spin-flip process dynamically polarizes the
nuclear spins, resulting in DPNS with a magnetization $\vec M$
\cite{gullan}. This spin-flip process mediates also an effective
spin transfer from electron spins to the nuclear spins.
Adopting the view that dynamical motion of DPNS is responsible to
the observed self-sustained current oscillation, we concentrate on 
the DPNS dynamics under the influence of both magnetic field and 
the above spin-flip process mediated spin transfer torque.
\begin{figure}
  \includegraphics[width=3.4 in]{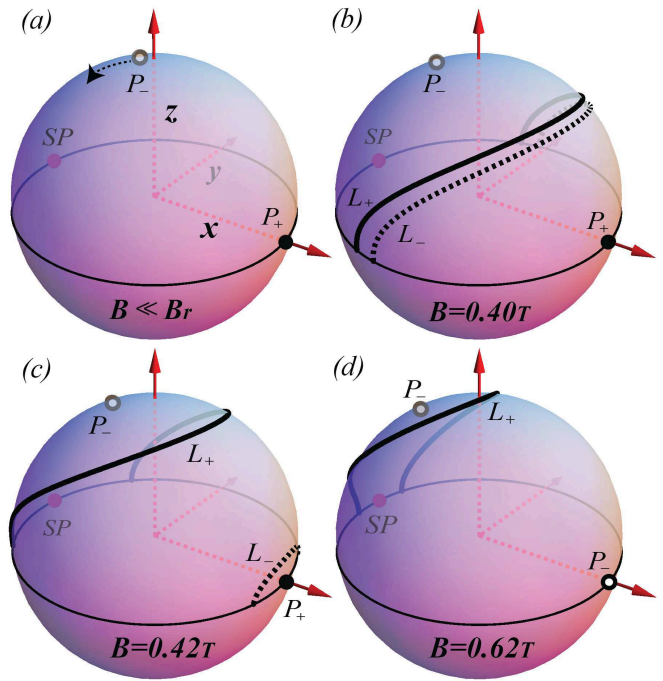}\\
\caption{(Color online) Change of attractors as an external
magnetic field varies: Only upper sphere was showed for clarity.
Solid (dash) black line $L_+$ ($L_-$) denotes stable (unstable)
limit-cycles (LCs). $P_+$ ($P_-$) labels stable (unstable) fixed 
points (FPs), and red dot SP is the saddle point at $\vec m=(-1,0,0)$.
(a) For $B\ll B_r$ and the only stable attractor is $P_+$;
(b) $B=0.40T$, a slightly over the semistable LC bifurcation where
$L_+$ and $L_-$ are generated; (c) $B=0.42T$, slightly below
subcritical Hopf bifurcation field at which $L_-$ merges with $P_+$
and becomes an unstable FP; (d) $B=0.62T$, slightly below the
homoclinic bifurcation field at which $L_+$ touches SP.}
\label{fig2}
\end{figure}

It is well-known that the Landau-Lifshitz-Gilbert equation governs 
the generic dynamics of a macro-spin preserving its magnitude while 
the so-called Landau-Lifshitze-Bloch equation is for the
dynamics of a macro-spin whose magnitude can also vary.
For simplicity and our belief that magnitude change of DPNS
magnetization is irrelevant to the observed oscillation, we assume
following dynamics for $\vec m=\vec M/{|\vec M|}$ \cite{xrw},
\begin{equation}
\frac{d\vec{m}}{dt}= -\vec{m}\times \vec {H}_{eff} +\alpha\vec{m}
\times\frac{d\vec{m}}{dt}+a\vec{m}\times(\vec{m}\times \hat{x}).
\label{LLG}\end{equation}
Here $t$ is in the units of $(\gamma M)^{-1}$, order of submicron
second for GaAs with $M=10^6 A/m$ and the nuclear
gyromagnetic ratio $\gamma=10(A\cdot s/m)^{-1}$ \cite{charnaya}.
The first term on the right-hand side of Eq. \eqref{LLG} describes the
Larmor precession around the effective field $\vec{H}_{eff}=\vec H-D
\vec m$ from both external magnetic field $\vec H$ along x-axis,
and demagnetic field $D\vec m$ (in the units of $|M|$).
Cylindrical disk-like dots in experiment \cite{tarucha} have
diagonal demagnetization factors with $D_x = D_y = 0$ and $D_z=1$.
In reality, a small difference between $D_x$ and $D_y$ exist either
due to the inhomogeneity or deviation of dots from the perfect
cylindrical shape, our numerical results show that the physics
reported here remain the same for $D_x \neq D_y$.
The second term is the phenomenological
Gilbert damping with a dimensionless constant $\alpha$.
The third term is the Slonczewski torque (per spin) along x-direction
\cite{slonc} originated from the transition from $|T^{-1}>$ to $|S>$
mentioned earlier (Fig. \ref{fig1}c). The Slonczewski coefficient
$a=\eta W/N$ is proportional to transition rate $W$ from $|T^{-1}>$
to $|S>$ and inversely proportional to the polarized nuclear number
$N\sim{10^5-10^6}$ \cite{tarucha}.
Dimensionless coefficient $\eta=S_{\perp}/\hbar$ measures the average
spin angular moment quanta transferred from one electron to nuclei.
$W$ in the Fermi golden rule approximation reads
\cite{fujisawa,macdonald,lopez}
\begin{equation}
W={W_0}\frac{\Gamma^2}{\Delta E^2 +\Gamma ^2}\times \xi
\label{bc}\end{equation}
Where \[\xi  = \left\{ {\begin{array}{*{20}{c}}
{1,\;\;\;\;\;\;\;\;\;\;\;\;\;\;\;}&{\;\;\;\;\Delta E > 0}\\
{\exp (\frac{\Delta E}{k_BT}),}&{\;\;\;\Delta E <0}
\end{array}} \right.\]
$\Delta E\equiv |g\mu_BB|-J$ is the level spacing between $|T^{-1}>$
and $|S>$ states. The effective Lande g-factor for GaAs is $g=-0.44$
\cite{yakobe2}. $k_B$ is the Boltzmann constant and $T=1.8K$ is the
experimental temperature. The level broadening $\Gamma$ of state
$|T^{-1}>$ is order of phonon energy of $\mu eV$ \cite{macdonald},
and ${W_{0}}\sim{10^3}-{10^4}$ \cite{tarucha2}
is the typical resonant spin flip-flop rate.

In the absence of the external field, all points on the equator of
$\vec m$-sphere are marginal stable fixed points (FPs) while 
$\vec m=(0,0,\pm 1)$ are unstable. Under a very small field, $\vec 
m=(-1,0,0)$ becomes the only saddle point (SP in Fig. \ref{fig2}) and 
$\vec m=(1,0,0)$ the only stable attractor ($P_+$ in Fig. \ref{fig2}). 
The two unstable FPs (upper $P_- $ in Fig. \ref{fig2}) move toward SP 
along the big cycle in $x-z$ plane as B increases (indicated by the 
arrow in Fig. \ref{fig2}a). In terms of energy, $\alpha$-term 
is always an energy sinker while $a$-term could be both energy 
sinker and source \cite{xrw}. In the vicinity of $P_+$, $\alpha$-term 
serves as an energy source that can be seen from the fact that 
$a$-term tends to push the system away from $P_+$. As $B$ approaching 
$B_r$, $a$-term may become large enough to destabilize $P_+$. 
This is confirmed by the standard stability analysis \cite{nbook} 
by calculating the Liapunov exponent at $P_+$.
The Liapunov exponents at $P_+$ become positive at about $B=0.42T$, 
and the system has no stable FPs at this point. This entails the 
existence of limit-cycle(s) (LCs) in two dimensions (current case). 
Indeed, our numerical calculations support this scenario. 
Fig. \ref{fig2} shows the locations of various attractors of Eq. 
\ref{LLG} at various $B$. For a field much smaller than $B_r$ 
(Fig. \ref{fig2}a), all phase flows end at $P_+$, the only stable 
FP of the system.
At a critical field $B_0$, slightly smaller than $0.40T$,
the system undergoes a semistable LC bifurcation (shown in Fig.
\ref{fig2}b at which a pair of LCs, one stable (solid black line)
and the other unstable (dashed line), appear simultaneously.
The two LCs move in opposite direction as the field increases further.
At a critical value $B=B_1$ around $0.42T$, the unstable LC merges
with $P_+$, undergoing a subcritical Hopf bifurcation and turning
$P_+$ into an unstable FP $P_-$. DPNS changes from a static state
to a LC state, the onset of self-sustained oscillation  (Fig. \ref{fig2}c).
The LC touches the saddle point at another critical value $B=B_2$ of
about $0.62T$, undergoing a homoclinic bifurcation (Fig. \ref{fig2}d).
The corresponding period diverges when $B$ goes to
$B_2$ because the phase velocity vanishes at the saddle point.
These are exactly what were observed in the experiment \cite{tarucha}.

\begin{figure}
  \includegraphics[width=3.4 in]{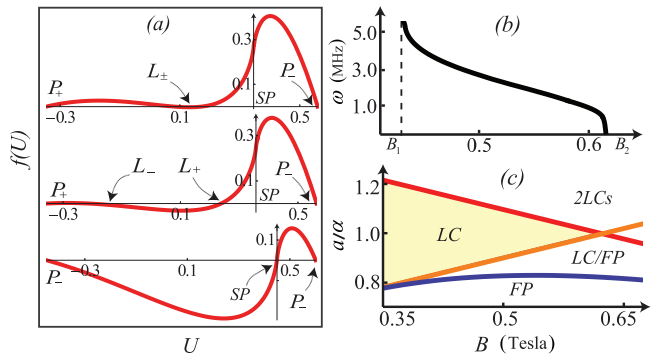}\\
\caption{(Color online) (a) The Melnicov curves $f(U)$ for $B=0.40T,\
0.42T,$ and $0.62T$ (from top to bottom). Zeros with positive (negative) slope are stable
(unstable) attractors. The far left node corresponds to $(1,0,0)$,
while the far right node  $(0,0,\pm 1)$ and the origin (crossing of
$x-$ and $y-$axis) was chosen to be the energy of SP.
(b) Field dependence of SSCO frequency $\omega$.
The period diverges at $B_2$.
(c) Phase diagram in $B- a/\alpha$ plane.
Red, orange, and blue curves are for homoclinic, subcritical Hopf,
and  semistable LC bifurcations, respectively.
FP, LC, LC/FP, and 2LCs denote fixed-point, one limit-cycle,
co-existence of limit-cycle and fixed-point, and two limit-cycles
phases, respectively.}\label{aaa}
\end{figure}

In order to obtain the field-dependence of period, one needs
to locate first the LC for a given field. This can be done by using
the Melnicov theory \cite{nbook,bertotti} valid for a nearly
conserved system of small $\alpha$ and $a$. It is well known that
the energy dissipation rate \cite{xr3}
\begin{equation}
f(U)\equiv \oint\limits_{{L(U)}} (-\alpha \frac{d\vec m}{dt}-a
\vec m \times \hat x)\cdot d\vec m
\end{equation}
is a function of equal energy contour $L(U)$, thus also a function
of energy $U\equiv \frac{1}{2}(D_xm_x^2 + D_ym_y^2+D_zm_z^2)-
\vec{H}\cdot\vec m$.
$f$ is the dissipated energy after the system moves along $L(U)$ once.
Interestingly, the Melnicov function of stable and unstable FPs and
LCs is zeros by definition. According to the Melnicov theory, the LC 
is approximated by one particular $L(U^*)$ that satisfies $f(U^*)=0.$
Fig. \ref{aaa}a is the U-dependence of f(U) for $B=0.40T, \ 0.42T,\
0.62T$ (from top to bottom) in the vicinity of $B_0$, $B_1$ and 
$B_2$ respectively.
By definition, the energy contours for energy extremes ($P_{\pm}$
minima and maxima) are points, thus the Melnicov function is zero there.
These correspond to the far left and far right nodes in the figure.
The origin (crossing of $x-$ and $y-$axis) was chosen to be the energy
of the saddle point. $f$-curve have positive (negative) slops are
stable (unstable). To see this, consider a node $U^*$ with
$\frac{\partial f}{\partial U}\left|_{U^*}\right. <0$ and a slightly
deviation of $U$ from $U^*$, say $U>U^*$, $U$ should decrease after
the system moving along the $L(U)$ once because $f(U)<f(U^*)=0$.
Thus the system tends to push $U$ back to $U^*$.
For $B=0.40T$, zero slop of $f(U)$ at the node
around $U=0.1$ corresponds to generation of a pair of LCs, leading
to a semistable LC bifurcation ($f(U^*)=0,\ \frac{\partial f}
{\partial U}\left| {_{U^*}} \right. = 0$). A slight increase of $B$
lowers $f(U)$ curve near $U^*=0.1$ and the node splits into two.
The left (right) one with negative (positive) slope corresponds to
stable (unstable) LC. The stable LC $L_+$ moves towards the SP
while unstable LC $L_-$ moves towards $P_+$ as $B$ increases.
The $L_-$ merges with $P_+$ around $0.42T$, and turns $P_+$
into an unstable FP $P_-$ (negative slop). This is a subcritical Hopf
bifurcation. Further increase of $B$ to $0.62T$, $L_+$ touches
saddle point SP and become a homoclinic loop, resulting in a
homoclinic bifurcation.
After locating the LC for a given field in the window of
$[B_1,B_2]$, the period of DPNS can be evaluated by
\begin{equation}
\tau=\oint\limits_{_{L(U^*)}} \frac{d \vec m}{\frac{d\vec m}{dt}}  ,
\label{ENERGYzero1}\end{equation}
where $d\vec m/dt$ is given by Eq.\eqref{LLG}.
As shown in  Fig. \ref{aaa}b, the period increases monotonically
with field and diverges at $B_2$.

% Discussions
The electron in triplet state ($|T^{-1}>$) plays dual roles.
It not only glues together the nuclear spins through the
hyperfine coupling so that nuclear spins become polarized,
but also generates a torque on the DPNS. The self-sustained current oscillation comes from the periodic motion of DPNS along an LC.
The LC originates from the instability of a static DPNS state
under two competing forces: One is energy input from the
tunneling electrons that drives the DPNS away from
its static state, a FP. The other is the dissipation of
$\alpha$-term that tends to push the DPNS to its FP.
The periodical motion of DPNS leads to the current oscillation. 
This theory explains why the current oscillation was only observed 
in the spin-blockaded regime and under an magnetic field.
It also explains why the period can be several order of
magnitude longer than the typical spin precession period.
In principle, the current oscillation reported here should exist 
in systems with more than two dots in the spin-blockaded regime. 
However, it may be easier to observe it in a spin-blockaded double 
quantum dots because it is easier to glue fewer nuclear spins in a 
smaller space. This is in a sharp contrast to the current oscillation 
observed in superlattices that originates from the negative 
differential resistance \cite{xrw}, where it is only observed in 
superlattices with more than 20 wells. It shall be interesting to 
see whether there are also field-induced Hopf and homoclinic 
bifurcations in Landau-Lifshitz-Bloch dynamics that governs spin 
dynamics whose magnitude can also vary.

Interestingly, if $a$ in Eq. \ref{LLG} can vary independently
from the field $B$, one can obtain the homoclinic bifurcation,
subcritical Hopf bifurcation, and  semistable LC bifurcation
curves as a function of $B$ and $a/\alpha$ in a similar way as 
what we explained earlier. Our results are showed in Fig. 3c. 
The phase diagram shows that Eq. \ref{LLG} supports various stable 
phases (stable attractors), including stable FPs only, coexistence 
of a stable FP and a stable LC, and existence of two stable LCs. 
In summary, there are a number of predictions in our theory to be 
confirmed. According to our analysis, a LC solution can only exist 
when spin transfer torque is large enough and self-sustained 
current oscillation appears and disappears when the tunneling current 
varies. The current can be controlled by the
electron coupling between electrodes and dots through gates.
Thus, this provides an experimental way of testing our theory.
Also, our theory predicts multiple stable attractors (Fig. 3c) in
certain parameter regions, either the coexistence of LC and FP or
the existence of two LCs. Thus one should expect hysteresis loops
or two oscillation periods if proper conditions are satisfied.

In conclusion, we showed that an external magnetic field can induce 
a Hopf bifurcation at a low field and a homoclinc bifurcation at 
a high field for DPNS in a spin-blockaded double quantum dots,
between which is the magnetic field window for the self-sustained 
current oscillation. The amplitude and period of the oscillation 
becomes bigger and bigger as the field increases, in good agreement 
with the experimental findings. The period diverges at the homoclinc 
bifurcation in our model which explains well why the period is of 
several orders of magnitude larger than the fundamental time scales. 

%-------------------------------------------------------------
This work is supported by Hong Kong RGC Grants (604109 and RPC11SC05).

\end{document}